\documentclass[aps,prb,twocolumn,showpacs]{revtex4}
\usepackage{graphicx}

\begin{document}
   \title{Vortex-vortex interaction in thin superconducting films}

\author{Ernst~Helmut~Brandt}
\affiliation{Max-Planck-Institut f\"ur Metallforschung,
    D-70506 Stuttgart, Germany}

\date{\today}

\begin{abstract}
The properties of vortices in superconducting thin films are
revisited. The interaction between two Pearl vortices in an
{\it infinite} film is approximated at all distances by a simple
expression. The interaction of a vortex with a regular lattice of
real or image vortices is given. The two spring constants are
calculated that one vortex in the vortex lattice feels when the
surrounding vortices are rigidly pinned or are free.
The modification of these London results by the
{\it finite size} of real films is obtained. In finite films, the
interaction force between two vortices is not a central
force but depends on both vortex positions, not only on their
distance. At the film edges the interaction {\it energy} is zero
and the interaction {\it force} is peaked. Even far from the
edges the vortex interaction considerably deviates from the Pearl
result and is always smaller than it.
\end{abstract}

\pacs{74.25.Qt,           
      74.78.Db, 74.78.Bz, 
      74.25.Ha}           

\maketitle

\section{Introduction}  

In this note the problem of vortices in thin superconducting films
is revisited. Useful formulae for the interaction between vortices
in thin superconducting films within London Theory are given when
the film thickness $d$ is smaller than the London penetration
depth $\lambda$. In such thin films the vortex interaction is
mediated mainly by the magnetic stray field, and the screening of
the in-plane supercurrents is governed by the effective
2-dimensional (2D) penetration depth
$\Lambda = \lambda^2 /d > \lambda > d$. The current density
$j(x,y,z)$ in thin films is nearly independent of the $z$
coordinate (perpendicular to the film) and the sheet current
in the film is thus ${\bf J}(x,y) =
 \int_{-d/2}^{d/2} {\bf j}(x,y,z) dz \approx{\bf j}d$.
The sheet current of a vortex in such a thin film and the
interaction force between two vortices was calculated first by
Judea Pearl \cite{1} for an infinitely large film. For infinite
films of arbitrary thickness the Pearl-London vortex and the
vortex-vortex interaction were calculated analytically,
\cite{2,3,4} and the Ginzburg-Landau (GL) theory for periodic
vortex lattices in such films with arbitrary GL parameter $\kappa$
in the entire range of inductions $\bar B$ was computed
in Ref.~\onlinecite{5}.

  For superconductors of finite size, recently there have been
numerous computations based on the static or time-dependent GL
equations for small (mesoscopic) specimens containing a few
vortices, also giant vortices with several flux quanta, and
considering the equilibrium state or the penetration, exit,
nucleation, and annihilation of vortices. For example,
superconducting disks are considered in
Ref.~\onlinecite{5a126,5a118,5a115117,5a78,5a61,5a34,5aZhou},
squares and other shapes in Ref.~\onlinecite{5b105,5b84,5b66,5b45},
also squares containing magnetic dots \cite{5c13,5c6988} or
antidots or blind holes \cite{5d84,5d66} as pinning
centers, and disks in inhomogeneous magnetic field.\cite{5e99}
When the 2D penetration depth
$\Lambda = \lambda^2 /d$ is much larger than the film size, the
energy of the magnetic stray field outside the specimen is expected
to be negligible and analytical approximations can be
made.\cite{5f71} Further work computes the properties
of infinitely long straight vortices in semi-infinite \cite{5gHern}
or cylindrical \cite{5hSard} samples and argues that these results
apply also to thin films in a perpendicular field in cases when the
magnetic stray-field energy outside the film may be disregarded.

  The present paper focusses on situations when the stray-field
energy is not negligible, e.g., since $\Lambda$ is smaller than
the film size, and it is restricted to the London limit in which
the superconducting coherence length $\xi$ and vortex core radius
$r_c \approx \xi$ are much smaller than the film size and than
$\Lambda$. In Sec.~II we first consider the limit of ideal
screening ($\Lambda = 0$) when the sheet current and energy of
vortices are completely determined by the stray field and can be
derived easily for an {\it infinite} film. Introduction of a finite
$\Lambda$ then leads to Pearl vortices, whose interaction is
discussed in some detail in Sec.~II-V and in Appendix A.
Finally, in Sec.~VI vortices and their interaction in films of
{\it finite size} are presented, computed by a method \cite{6,7}
that applies to thin flat films of any size and shape and to any
$\Lambda$. As a main result in Sec.~VII it will be shown that
the sheet current and interaction of vortices in films of finite
size differ considerably from the ideal Pearl result. As expected,
this deviation is most pronounced near the film edges, where
the real interaction {\it potential} vanishes while the
interaction {\it force} is peaked. But less expectedly, this
deviation is considerable even when the interacting vortex
pair is located far from the film edges, e.g., near the
center of a square film: The real interaction with
a central vortex is reduced from its Pearl value by a factor
which decreases approximately linearly from unity to zero as
one goes from the central vortex towards the edge, see
Fig.~6 below. A summary is given in Sec.~VIII.

\section{Vortices in thin infinite films}  

The thin-film problem differs from the behavior of currents
and vortices in bulk superconductors by the dominating role
of the magnetic stray field outside the film. The interaction
between vortices occurs mainly by this stray field and thus
has very long range (interacting over the entire film width),
while in bulk superconductors the vortex currents and the
vortex interaction are screened and thus decrease exponentially
over the length $\lambda$.

The long-range currents and forces can be shown for the ideal
screening limit of zero $\lambda$ as follows. Consider one vortex
in the center of a large circular film with radius $R \to \infty$.
On length scales larger than $\lambda$ this point vortex behaves
like a magnetic dipole, composed of two magnetic monopoles: one
sends magnetic flux $\Phi_0$ into the space above the film and
one receives the same flux from the lower half space. Here
$\Phi_0 = h/2e = 2.07 \cdot 10^{-15}$ Tm$^2$ is the quantum of
magnetic flux. The magnetic field lines of this point-vortex are
straight radial lines, all passing through this point. The
magnitude of this magnetic stray field is
$\Phi_0 /(2\pi r_3^2)$ above and $-\Phi_0 /(2\pi r_3^2)$ below
the film, since the flux $\Phi_0$ has to pass through the shell
of any half sphere of 3D radius $r_3 = (r^2 + z^2)^{1/2}$ and
surface area $2\pi r_3^2$. Here $r = (x^2 +y^2)^{1/2}$ is the
2D radius in the film plane $z=0$. Directly
above and below the film plane the stray field has the same
magnitude $\Phi_0/(2\pi r^2)$ but has opposite sign. This jump of
the magnetic field component parallel to the film is caused by a
sheet current that circulates around the vortex and equals this
field difference in size, $J(r) = \Phi_0 /(\mu_0 \pi r^2)$. This
current thus decreases very slowly with radius $r$. The  force $F$
between this central vortex and a second vortex at a distance $r$
is radial (central force) and repulsive and in size equals just
this sheet current $J(r)$ times the flux quantum $\Phi_0$,
  \begin{eqnarray}  
  F(r) =\Phi_0 J(r) ={\Phi_0^2 \over \mu_0\, \pi\,r^2}.
  \end{eqnarray}
  The interaction potential $V(r)$ between these two
point vortices (Pearl vortices) follows from its derivative
$F(r) = -V(r)'$ and the definition $V(\infty) =0$,
  \begin{eqnarray}  
  V(r) = {\Phi_0^2 \over \mu_0\, \pi\,r}.
  \end{eqnarray}
Note that this result does not depend on
$\Lambda = \lambda^2 /d$.  It applies also for finite
$\Lambda$ if the distance is large, $r \gg \Lambda$.

  The general results for arbitrary $\Lambda$ can be
derived from the expression for the interaction potential
noting that still $F(r) = -V(r)'$ and $J(r) = F(r)/\Phi_0$.
One has
  \begin{eqnarray}  
  V(r) = {\Phi_0^2 \over \mu_0} \int\!{d^2 k \over 4\pi^2}
    {2 \cos {\bf k r} \over k + 2\Lambda k^2}
  = {\Phi_0^2 \over \mu_0} \int_0^\infty\!\! {dk \over 2\pi}
    {2 J_0(k r) \over 1 + 2\Lambda k} \,,
  \end{eqnarray}
where $J_0(x)$ is a Bessel function and $k^2 = k_x^2+k_y^2$.
The limits for small
and large distances are, \cite{1} see also \ Eq.~(2),
  \begin{eqnarray}  
  V(r) &\approx& {\Phi_0^2 \over \mu_0}~ {\ln(2.27\Lambda/ r)
    \over 2\pi \Lambda} ~\,~{\rm for~~} r \ll \Lambda\,, \\
  V(r) &\approx& {\Phi_0^2 \over \mu_0}~ {1 \over \pi r}~~~
              ~~~~~~~~~~~~{\rm for~~} r \gg \Lambda\,.
  \end{eqnarray}
The factor 2.27 is a fitted constant that follows from
the numerical evaluation of the integral (3).
An excellent fit to the exact result valid for all distances
$0 < r < \infty$ and all $\Lambda$ is within line thickness
  \begin{eqnarray}  
  V(r) \approx {\Phi_0^2 \over 2 \pi \Lambda\mu_0}~ \ln\Bigg(\,
   {2.27\Lambda\over r} -{0.27\Lambda\over 9\Lambda+r} +1 \Bigg) .
  \end{eqnarray}
Figure 1 shows the exact potential from Eq.~(3), its
limits (4) and (5), and the approximation (6). One can see
that the expression (6) (dots) practically coincides
with the exact potential (solid line).

In the original paper by Pearl \cite{1} the force
$f_{12}(r) = - V'(r)$ between two point vortices  was
expressed in terms of  Struve and Neumann functions
$S_1$ (or ${\bf H}_n$) and $N_1$ (or $Y_n$, the Bessel
function of the second kind, or Weber function),\cite{8}
  \begin{eqnarray}  
  f_{12}(r) = {\Phi_0^2 \over 8 \Lambda^2 \mu_0}~
  \Big[\,{\bf H}_1\Big( {r\over 2\Lambda} \Big)
    - Y_1\Big( {r\over 2\Lambda} \Big)-{2\over\pi} \Big] \,.
  \end{eqnarray}
Though both $S_1$ and $N_1$ are oscillating functions this
force $f_{12}$ is monotonic and agrees with $-V'(r)$, Eq.~(3).
The interaction potential $V(r)$ may be also expressed in terms
of Struve and Weber functions, see appendix in \cite{9},
  \begin{eqnarray}  
  V(r) = {\Phi_0^2 \over 4 \Lambda \mu_0}~
  \Big[\,{\bf H}_0\Big( {r\over 2\Lambda} \Big)
        - Y_0\Big( {r\over 2\Lambda} \Big) \Big] \,.
  \end{eqnarray}

 \begin{figure}[bht]  
\includegraphics[scale=0.5]{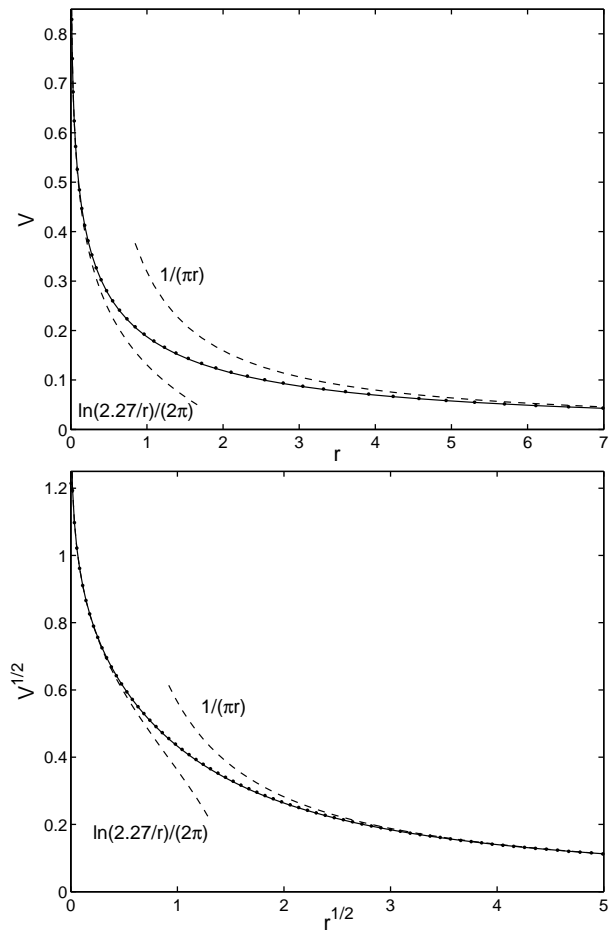}
\caption{\label{fig1}
The interaction potential $V(r)$, Eq.~(3), between two
Pearl vortices (solid lines), the limits (4) and (5)
(dashed lines) and the approximation (6) (dots). $V$ is
in units $\Phi_0^2 /(\mu_0 \Lambda)$ and $r$ in units
$\Lambda = \lambda^2/d$. The upper plot shows directly
$V(r)$. The lower plot shows $\sqrt V$ versus $\sqrt r$
to depict larger ranges of $V$ and $r$ and to show the
validity of the limiting expressions at small and large $r$
more clearly.}
 \end{figure}  

\section{Interaction with vortex lattices}  

Next I consider the interaction potential $V({\bf r})$
between a vortex at position ${\bf r} = (x,y)$ in the film
and a regular lattice of vortices sitting at ideal-lattice
points ${\bf R}$ with reciprocal lattice vectors ${\bf K}$.
This situation can be interpreted in two ways:

1. This lattice can consist of real vortices in the film,
generated by a constant applied magnetic field $B_a$
perpendicular to the film plane (along $z$). In this
case it will be typically a triangular lattice with
density $n = B_a/\Phi_0$, since the magnetic field has
to penetrate the infinitely extended film completely,
i.e., the average induction $\bar B$ equals $B_a$.
The lattice spacing is then
$a \approx (\Phi_0 / \bar B)^{1/2}$.

2. In a second application the lattice may be formed
by the image vortices sitting on a superlattice that is
constructed to achieve periodic boundary conditions in the
numerical simulation of a film with vortex pinning. In this
case the superlattice may be chosen such that the basic cell
is a rectangle with side lengths ($L_x, L_y$). The
regular lattice positions are then
${\bf R} = (\mu L_x, \nu L_y)$ where $\mu, \nu = 0, \pm1,
\pm2, \dots $ are integers, and the reciprocal lattice
vectors are ${\bf K} = (2\pi\mu /L_x, 2\pi\nu /L_y)$.

  The interaction potential between the probing vortex at
position ${\bf r}$ and the periodic lattice of vortices
(or images) is obtained by linear superposition of
potentials of the form (3),
  \begin{eqnarray}  
  V_{\rm per}({\bf r}) = {\Phi_0^2 \over \mu_0} \int\!{d^2 k
  \over 4\pi^2}  \sum_{\bf R} {2 \exp[i {\bf k (r-R)}]
  \over k + 2\Lambda k^2} \,.
  \end{eqnarray}
This potential is periodic,
$ V_{\rm per}({\bf r+R}) =V_{\rm per}({\bf r}) $.
The infinite lattice sum in (8) can be evaluated using
the formula
  \begin{eqnarray}  
  \sum_{\bf R} \exp(i {\bf k R}) = 4\pi^2 n
  \sum_{\bf K} \delta_2({\bf k-K}) \,,
  \end{eqnarray}
where $n$ is the density of lattice points ${\bf R}$;
the density of reciprocal lattice points is
$1/(4\pi^2 n)$. Inserting (10) into (9) and performing
the ${\bf k}$ integration over the 2D delta functions
$\delta_2({\bf k-K})$ one obtains
  \begin{eqnarray}  
  V_{\rm per}({\bf r})={\Phi_0^2\,n\over\mu_0} \sum_{{\bf K}\ne0}
  {2 \cos({\bf Kr}) \over K + 2\Lambda K^2} \,.
  \end{eqnarray}
This periodic potential may be used for computer
simulation of vortex pinning in a thin film with
periodic boundary conditions. In this case the
super-lattice density is $n=1/(L_x L_y)$.

  Next I apply this potential to the following problem.
Consider a regular (e.g. triangular) vortex lattice in
the film in which all vortices are pinned to these ideal
vortex positions and the central vortex at ${\bf R}=0$
is moved a bit in $x$ direction. Which interaction
potential with all other vortices does this central
vortex see? To obtain this potential $V_c({\bf r})$, or its
curvature $V_c''(0)=\partial^2 V_c({\bf r}) / \partial x^2$
at ${\bf r} = 0$, we have to subtract from the complete
lattice sum (9) the term with ${\bf R}=0$, since this
term describes the interaction of the shifted vortex
at ${\bf r} \approx 0$ with the unshifted vortex at
${\bf R}=0$, which would be very large and diverging as
${\bf r \to R}$. The resulting potential is
 \begin{eqnarray}  
 V_c({\bf r}) = {\Phi_0^2 \over \mu_0} \int\!{d^2 k
  \over 4\pi^2}  \sum_{{\bf R} \ne 0} {2 \exp[i {\bf
   k (r-R)}] \over k + 2\Lambda k^2} ~~~~\nonumber\\
 = {\Phi_0 \bar B \over \mu_0} \Big[ \sum_{{\bf K}\ne0}
  {2 \cos({\bf Kr}) \over K + 2\Lambda K^2}
 -\! \int\!{d^2 k \over 4\pi^2 n}\, {2 \cos({\bf kr})
  \over k + 2\Lambda k^2} \Big] \,.
 \end{eqnarray}
with $\bar B = n \Phi_0$.
This result is still general and applies to any ${\bf r}$.
To obtain its curvature at ${\bf r}=0$ we apply to it the
Laplace operator $\nabla^2 = \partial^2 / \partial x^2 +
\partial^2 / \partial y^2$. This creates a factor $-K^2$
in the sum and a factor $-k^2$ in the integral.
Due to symmetry of the
triangular (and also the square) vortex lattice, the
curvature of $V_c({\bf r})$ at ${\bf r}=0$ is the same along
all directions, i.e., one has  $V_c({\bf r}) = V_c(0) +
 {1\over 2} V_c''(0) r^2 + O(x^4, x^2 y^2, y^4)$ with
 $ V_c''(0) = \partial^2 V_c /\partial x^2 =
 \partial^2 V_c /\partial y^2 = {1\over 2} \nabla^2 V_c$
 at $x=y=0$. The result for this curvature is thus
  \begin{eqnarray}  
  V_c''(0) = {\Phi_0 \bar B \over 2 \mu_0} \Big[ \!
   \int\!\! {d^2 k \over 4\pi^2 n}\, { 2k^2
   \over k\! + 2\Lambda k^2} - \! \sum_{{\bf K}\ne 0}
   {2 K^2 \over K\!\! +\! 2\Lambda K^2} \Big] .\,
  \end{eqnarray}
The sum and the integral in (13) largely compensate.
This may be seen by formally introducing an upper
boundary $k_{\rm max} = K_{\rm max}$, which makes the
integral and the sum finite but drops out from their
difference. The main contribution to this difference
comes from the integral at small $k \le k_B$, where
$k_B$ is the radius of the Brillouin zone
approximated by a circle. One has $k_B^2 = 4\pi n$
for all vortex lattice symmetries.

  For $\Lambda$
larger than the vortex spacing $a$, one has
$\Lambda K \gg 1$ for all $K$. More precisely, the
condition is $K_{10}^2 = 16\pi^2/(3a^2) \gg \Lambda^{-2}$,
thus $2\pi \Lambda \gg a$ is sufficient. In this case the
general result (13) may be evaluated approximately by
keeping just the integral over $0 < k \le k_B$ and
noting that the remaining integral for $k>k_B$ is
compensated by the sum over ${\bf K} \ne 0$. The
final result for the curvature in this limit is then
 \begin{eqnarray}  
  V_{c,\infty}''(0)
  = {\Phi_0 \bar B \over 2 \mu_0 \Lambda}
  = {\Phi_0 \bar B d \over 2 \mu_0 \lambda^2}\,.
 \end{eqnarray}

\section{Results from Elasticity}  

The spring constant (14) may be calculated also from the
elasticity theory of the vortex
lattice, \cite{10,11,12} see Appendix A. In this case one
requires nonlocal elasticity with a dispersive compressional
modulus $c_{11}(k)$. The result coincides with Eq.~(14).

In Appendix A also is calculated the spring constant
$V''_{\rm elast}(0)$ felt by the central vortex when the
surrounding vortices are not pinned but are held in place
only by the interaction with their neighbors. In this case,
the elastic response of the central vortex depends only
on the shear modulus $c_{66}$ of the triangular vortex
lattice, \cite{10,12} which is not dispersive.
Thus, local elasticity theory is sufficient for this problem.
However, now an inner cut-off radius $k_{\rm min} = 1/R_d$ is
required for the ${\bf k}$ integral, which is taken from the
finite radius $R_d$ of the thin disk containing the vortex
lattice. It is assumed that the vortices are pinned at the
edge of the disk, at $r=R_d$. The resulting elastic spring
constant is then
  \begin{eqnarray}  
  V''_{\rm elast}(0) ={\bar B \, \Phi_0 \over
  4 \mu_0 \Lambda \ln(R_d k_B)}
  ={V''_{c,\infty}(0) \over2 \ln(\pi R_d/a)}\,.
  \end{eqnarray}
This  $V''_{\rm elast}(0)$  is smaller than the spring constant
$V''_{c,\infty}(0)$ of the rigidly pinned vortex lattice by a
factor $0.5 / \ln(\pi R_d/a )$. A constant force $f_0$ on the
central vortex thus displaces this vortex a distance $u_0$
that increases by a factor $2 \ln(\pi R_d/a )$ when pinning of
the surrounding vortices is switched off.

In all the above expressions the vortex core radius
$r_c \approx \xi$ and the coherence length $\xi$ were
assumed to be smaller than any other relevant length,
in particular $\xi \ll \Lambda = \lambda^2/d$. This
is the London limit. Finite $\xi$ is easily considered
by taking the $k$ integral not to infinite $k$ but to
some upper cut-off value $k_{\rm max} \approx 1/\xi$.
This cut-off removes the logarithmic divergence of
$V(r) \propto \ln(2.27 \Lambda/r)$, Eq.~(4),
and smoothes it over the core radius $r_c \approx \xi$.
This smoothing may also be performed by replacing in
the logarithm $r$ by $(r^2 + r_c^2)^{1/2}$ as shown
in Ref.~\onlinecite{13,14,15}.
The exact numerical solution of the GL equations for
a vortex and the periodic vortex lattice in films of
finite thickness is given in Ref.~\onlinecite{5}. The London
solution for a vortex in films of finite thickness is
presented e.g. in Ref.~\onlinecite{4}.

\begin{figure}[bht]  
\includegraphics[scale=0.5]{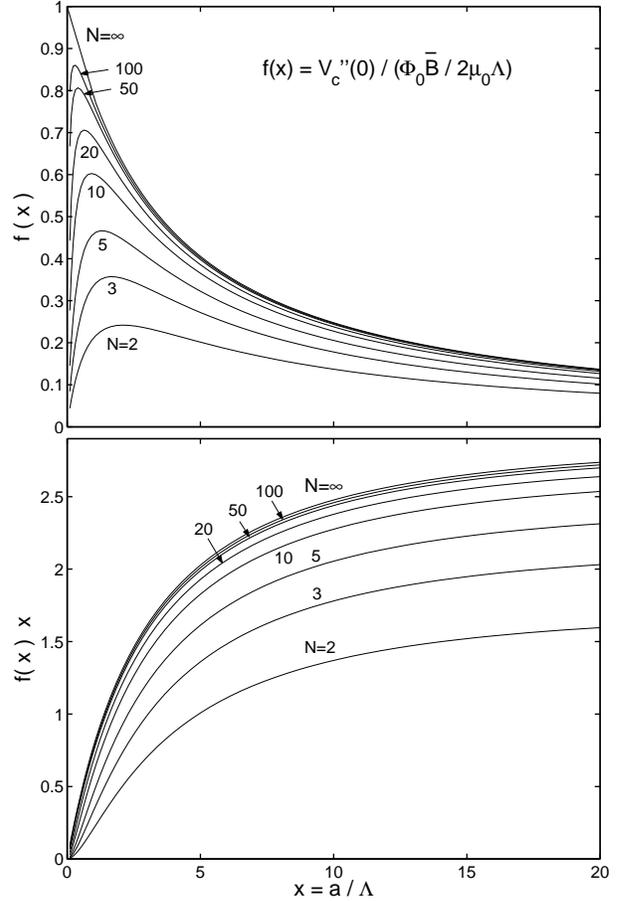}
\caption{\label{fig2}
The reduced curvature
$f(x,N) = V''_c(0,N)\cdot (2\mu_0 \Lambda/\Phi_0 \bar B)$,
Eq.~(17), of the potential exerted on a central Pearl vortex
by the surrounding $\pi N^2 a^2$ pinned vortices within a
circular area of radius $Na$.
The function $f$ is plotted versus the reduced vortex
spacing $x = a/\Lambda$ for several values of $N = 2$ to
$N =\infty$. While $f(x,N)$ (upper plot) for finite $N$ has a
maximum and vanishes at small and large $x$, the product
$f(x,N) \cdot x$  for all $N$ is monotonically increasing
with $x$.  }
\end{figure}  

\begin{figure}[bht]  
\includegraphics[scale=0.6]{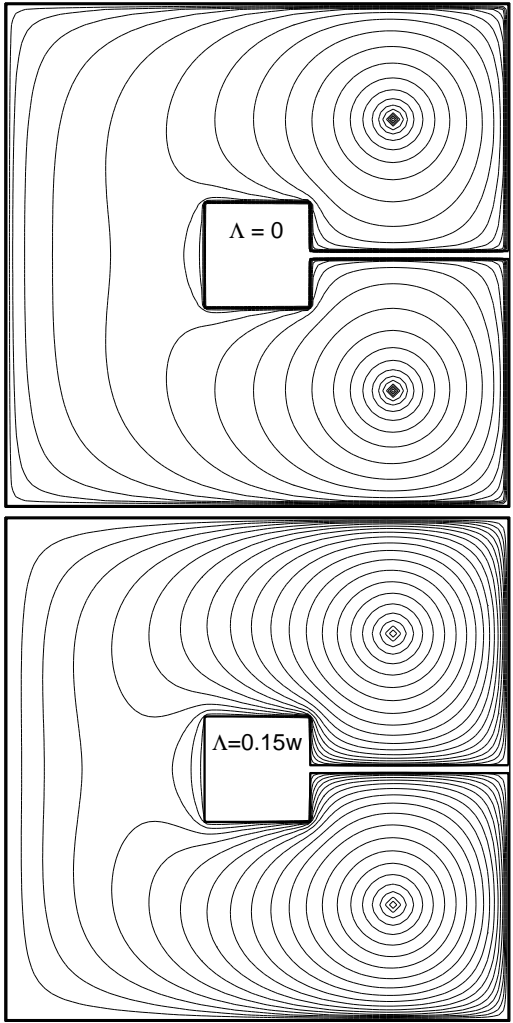}
\caption{\label{fig3}
Current stream lines of  a pair of vortices, coinciding with
the contour lines of the vortex-vortex interaction. Shown are
the contours of the logarithm of the interaction potential,
Eq.~(18), for a square thin superconducting film of size
$2w \times 2w$ with slit and central square hole, containing
a pair of Pearl vortices (due to the mirror symmetry of this
computation) at $x_i=0.54 w$, $y_i = \pm x_i$; $88\times 88$
grid points are used and 20 contours.  The force exerted by
this vortex pair on a third probing vortex at position $x,y$
acts perpendicular to these contour lines.
Top: 2D penetration depth $\Lambda = 0$
(ideal screening). Bottom: $\Lambda = 0.15 w$.}
\end{figure}  

\begin{figure}[bht]  
\includegraphics[scale=0.5]{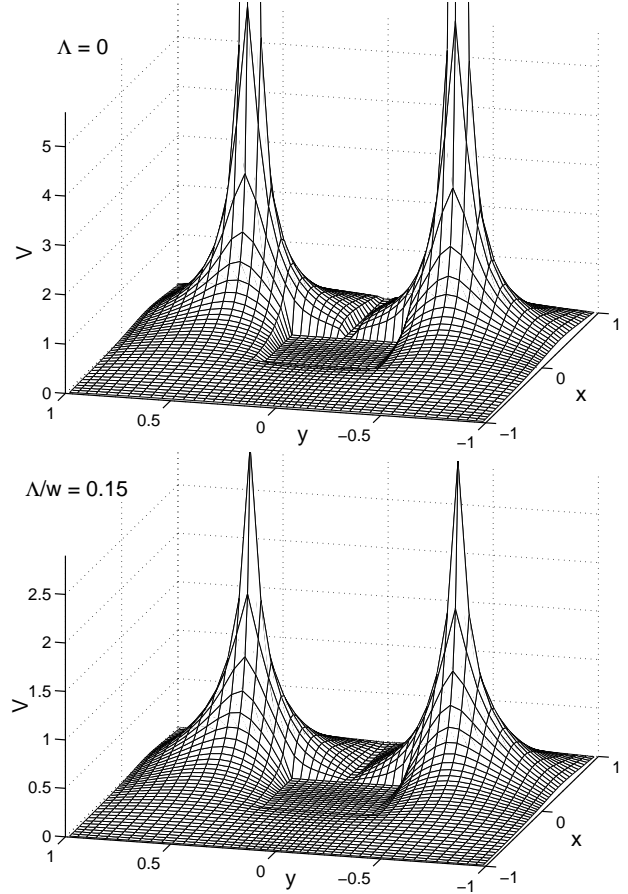}
\caption{\label{fig4}
3D plot of the interaction of a probing Pearl vortex
positioned at $(x,y)$ with a pair of vortices sitting
at $x_i=0.4 w$, $y_i = \pm 0.5 w$ in a thin square film
of size $2w \times 2w$ with slit and hole. Plotted is the
interaction potential, Eq.~(18), in units $\Phi_0^2/(\mu_0 w)$,
versus $x,y$ in units of $w$; $46\times 46$ grid points
are used. Top: $\Lambda = 0$ (ideal screening).
Bottom: $\Lambda = 0.15 w$.}
\end{figure}  

\begin{figure}[bht]  
\includegraphics[scale=0.5]{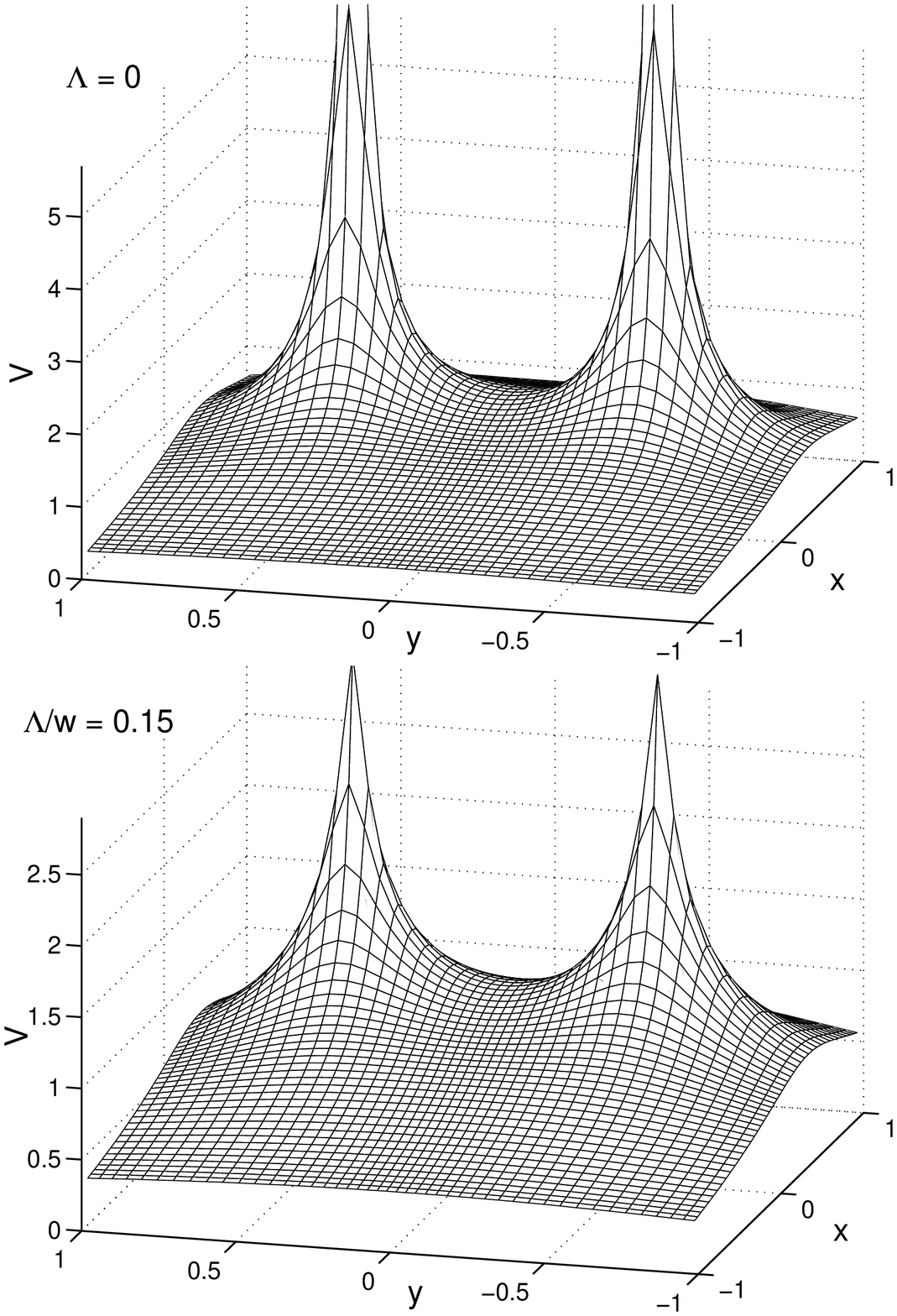}
\caption{\label{fig5}
Same geometry as Fig.~4, but Pearl potential, Eq.~(19).
This approximate interaction potential between a probing
Pearl vortex at $(x,y)$ with a pair of vortices
at $x_i=0.4 w$, $y_i = \pm 0.5 w$ does not vanish at the
edges of the square film and of its slit and hole, while
the correct numerical interaction (18) depicted in Fig.~4
does vanish there.}
\end{figure}  

\section{Vortex lattice of finite size}  

Expressions (12) to (14) apply to an infinitely extended
vortex lattice. The finite size of the superconducting
film can be approximately accounted for by taking the
sums $\sum_{{\bf R} \ne 0}$ in Eq.~(12) over a finite
area only, e.g., over a circular area of radius
$R_d = Na$ (disk radius), $|{\bf R}| < R_d$, containing
$\approx \pi N^2$ vortices. The approximate potential
that a central vortex sees in the presence of a regular
lattice of pinned vortices in such a thin disk is thus
 \begin{eqnarray}  
  V_c({\bf r}, N) = \sum_{\bf R} V({\bf r-R}),
  ~~ 0 < |{\bf R}| < Na  \,.
 \end{eqnarray}
For $N \to \infty$, Eq.~(16) coincides with Eq.~(12), and
the curvatures coincide, $V_c''(0,\infty) = V_c''(0)$,
Eq.~(13). For finite $N$, the interaction and its curvature
are reduced. To visualize this we define the dimensionless
curvature
 \begin{eqnarray}  
  f(a/\Lambda, N) = V_c''(0, N) / V''_{c,\infty}(0)\,,
 \end{eqnarray}
with $f(0, \infty) = 1$. This function is
shown in Fig.~2. One can see that the dense-lattice
limit, Eq.~(14), is a good approximation when
$a \le 2\Lambda$ and $N \ge 20$. For larger $a$ and smaller
$N$ the reduced curvature $f(a/\Lambda, N)$ decreases.
At small values of $x=a/\Lambda \le 2$ for $N \ge 2$,
$f(x)$ has a maximum and decreases to zero when $x\to 0$
because then also the disk radius $R_d =aN$ goes to zero.

\section{Vortex interaction in films of finite size} 

  The interaction potential and force between two vortices
in thin films of finite size and arbitrary shape depends
on the size and shape of the film. Moreover, it depends on
the positions of both vortices,
$V = V({\bf r}_1, {\bf r}_2)$, while for infinite films
(Sec.~I) $V(r)$ depends only on the distance
$r = |{\bf r}_1- {\bf r}_2|$. These dependences are
obvious from the fact that the interaction potential between
two vortices has to vanish when one vortex approaches the
edge of the film, where its magnetic flux goes to zero
since supercurrents cannot circulate around a vortex core
positioned on the edge. Furthermore, the interaction force
in general is {\it not a central force} but close to the film
edge the force on the second vortex is directed perpendicular
to the edge no matter where the first vortex is positioned.
This is so since the current generated by a vortex flows
along the film edge close to the edge, see Fig.~3.
Though the vortex interaction energy vanishes at the edges,
the interaction force has a peak there which increases with
decreasing $\Lambda$. Only when the two vortices are far
from all film edges and are close to each other, then the
central Pearl interaction potential $V(r)$ of Sec.~I is a
good approximation, see Sec.~VII.

  The vortex interaction in realistic films of finite extension
can be computed by the method described in Ref.~\onlinecite{6,7}.
One has
$V({\bf r}_i,{\bf r}_j) =V_{ij} =V_{ji} =-\Phi_0 g_i({\bf r}_j)$
where $g_i({\bf r})$ is the stream function of the 2D sheet
current density ${\bf J}_i({\bf r})$ caused by a vortex
centered at ${\bf r}_i$. In general one has
${\bf J(r)} = -{\bf\hat z} \times \nabla g({\bf r}) =
 \nabla \times ({\bf\hat z}g) = ( \partial g / \partial y,\,
 -\partial g / \partial x )$.
The function $g({\bf r})=g(x,y)$ has several useful properties
listed in Sec.~2A of Ref.~\onlinecite{7}. In particular, its
contour lines are the stream lines of the sheet current,
and it may be put $g=0$ on the outer edge of the film, which
coincides with a stream line if there is no current fed in
by contacts.

   The function $g(x,y)$ and its sheet current
${\bf J}(x,y)$ can be caused by an applied magnetic field,
or by the flux trapped in a hole in the film, or by vortices,
and also by applied currents, which we do not consider here.
Within London theory all these contributions superimpose
linearly. In Ref.~\onlinecite{7} it is shown how these sheet
currents can be computed by introducing a grid of $M$
equidistant or nonequidistant points ${\bf r}_i$ and then
inverting a $M\times M$ matrix. The inverted matrix
$K_{ij}^\Lambda$, Eq.~(11) of Ref.~\onlinecite{7},
(dimension m) is
closely related to the vortex interaction. One has
   \begin{eqnarray}  
   V({\bf r}_i, {\bf r}_j) = V({\bf r}_j, {\bf r}_i)
   = - \mu_0^{-1}  \,\Phi_0^2\,  K_{ij}^\Lambda /w_j \,,
   \end{eqnarray}
where $w_i$ are the weights of the grid points (dimension m$^2$).
This potential is repulsive (positive) and sharply peaked at
${\bf r}_i ={\bf r}_j$, and like the function $g(x,y)$
it vanishes at the outer edge of the film. This vanishing
is linear for $\Lambda > 0$, while for $\Lambda = 0$
(ideal screening) $g$ and $V$ go to zero like $\sqrt{\delta}$
where $\delta$ is the distance to the edge.

  If the film contains one or more holes (is multiply
connected) the $g(x,y)$ can have a non-zero value inside the
hole, $g(x,y) = I =$ const, where $I$ is the total current
that circles  around the hole when the hole contains trapped
magnetic flux. If the hole is connected with the outer edge
by a slit, then its border is part of the outer edge and
one has $g=0$ along the entire (inner and outer) edge of the
film. If the slit is short-cut, one can trap flux in the slit
and hole. If the slit is bridged by a superconducting weak link,
this film can be used as a Superconducting Quantum
Interference Device (SQUID). \cite{7,16}

Figure 3 shows the contour lines of this vortex interaction,
coinciding with the stream lines of the sheet current
${\bf J}(x,y)$ since $V \propto g$. Depicted are the contours
of the logarithm of the interaction potential, Eq.~(18),
between a probing vortex sitting at $(x,y)$ and a pair
of vortices  at ($x_i, \pm y_i$) since our
computation assumes mirror symmetry about the $x$ axis,
$\ln [ V(x_i,y_i;\,x,y) + V(x_i,-y_i;\,x,y) ]$.
In the regions where $V$ is very small, this logarithm shows
more contours than a linear contour plot would show.

Figure 4 shows the same interaction depicted as a 3D plot.
Both Fig.~3 and Fig.~4 show the geometry of a thin
superconducting square of size $2w\times 2w$ with a central
square hole and an open slit, which may be used as a SQUID.
Depicted are the cases $\Lambda = 0$ (ideal screening)
and $\Lambda/w = 0.15$. One can see that for $\Lambda = 0$
the potential $V$ and the stream function $g$ at the edges
go to zero with vertical slope,
$V \propto g \propto \sqrt{\delta}$ ($\delta =$ distance
to the edge), and for finite $\Lambda$,
 $V \propto g \propto \delta$ vanish linearly at the
inner and outer edges of the film. In Fig.~4, a grid of only
 $46 \times 46$ nonequidistant points is chosen such that no
grid points sit directly on the film edges. Therefore,
on the grid points closest to the outer edges of the film the
depicted $V$ is not exactly zero even when it is zero on the
very edges.

  Figure~5 shows, for the same geometry and grid as in
Fig.~4, the approximate vortex-vortex interaction obtained
from the Pearl potential of Sec.~1 valid for an infinite
film. To allow comparison with the correct numerical
potential of Fig.~4, which has mirror symmetry, we
symmetricize this approximate potential by plotting
  \begin{eqnarray}  
  V({\bf r}, {\bf r}_i) &=&
  V\big[\big((x-x_i)^2+(y-y_i)^2+\epsilon\big)^{1/2}\big]
              \nonumber \\  &+&
  V\big[\big((x-x_i)^2+(y+y_i)^2+\epsilon\big)^{1/2}\big]\,,
  \end{eqnarray}
where $V(r)$ is the Pearl interaction of Sec.~I, e.g., Eq.~(6).
The cut-off $\epsilon =4\cdot 10^{-4} w^2$ was chosen to reach
the same peak heights as in Fig.~4. Note that the numerical
interaction potential, Eq.~(18) and Fig.~4, has a natural
cut-off and finite peak heights, which are related to
the spacing of the grid points. The peaks of the numerical $V$
(Fig.~4) are well approximated by the cut-off Pearl peaks of
Eq.~(19) (Fig.~5) if they are far from all edges.

  As the peak position ${\bf r}_i$ of $V({\bf r, r}_i)$
approaches the film edge, the amplitude of the correct
potential and its peak height decrease and finally vanish as the
edge is reached. This means that the interaction {\it energy}
of two vortices is strongly reduced when both vortices are
close to the film edge. The interaction {\it force}, however,
i.e., the slope of $V$, may still be large, especially when
 $\Lambda$ is small and ${\bf r}$ or ${\bf r}_i$ or both
are close to the edge.

The $\sqrt\delta$ behavior of the vortex interaction near
the film edge can be seen in Fig.~5 of Ref.~\onlinecite{17},
which shows this potential (there called integral kernel
$\bar K$) for an infinite (along $y$) thin strip with
$\Lambda =0$, containing a dense row of vortices (same $x_i$,
many $y_i$); also shown there is the product
$\bar K(x, x_i)\cdot |x-x_i|^3$ that enlarges the
$\sqrt\delta$ shape at the edges. This ideal-screening potential
has a $\ln|x -x_i|$ singularity obtained by integrating the
$1/|{\bf r -r}_i|$ singularities of the Pearl potential
(for $\Lambda =0$) along the positions $y_i$.

\begin{figure}[bht]  
\includegraphics[scale=0.49]{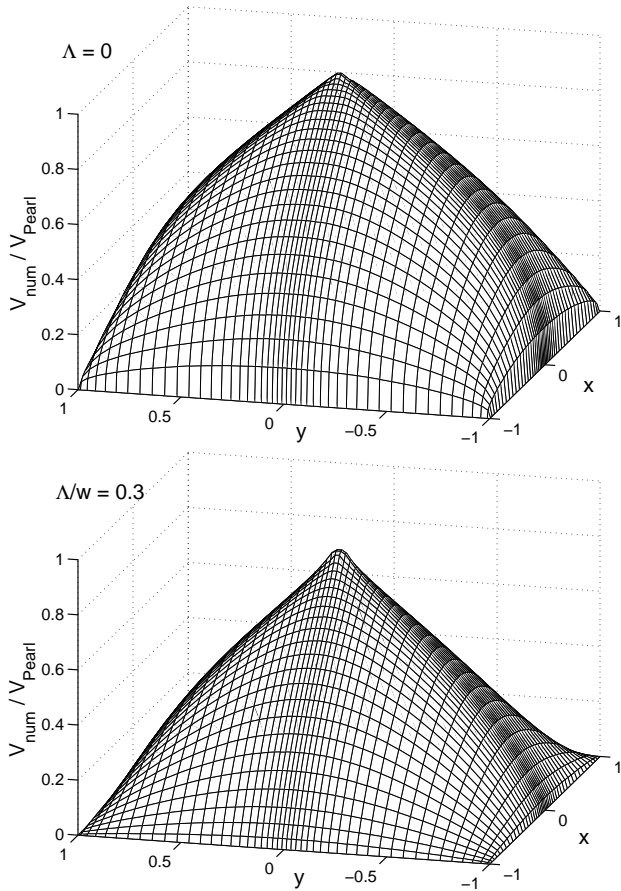}
\caption{\label{fig6}
The ratio of the exact numerical interaction potential
between two vortices and the Pearl interaction,
$V_{\rm num}(x,y) / V_{\rm Pearl}(x,y)$, for a thin
square of size $2w \times 2w$ with one vortex positioned
at the square center ($0,0$) and the other vortex at
$(x,y)$. Top: $\Lambda =0$. Bottom: $\Lambda = 0.3 w$.}
\end{figure}  

 \section{Ratio of numerical and Pearl potentials}   

  The analytical Pearl potential (19) (Fig.~5) does not depend on
the shape of the film, in particular, it does not vanish at the
outer and inner edges as it should. If an approximate analytic
interaction potential is needed, e.g., for computer simulation
of vortex motion and vortex pinning in films, one may
construct this for simple film shapes like circular disks or
rectangles or long strips, by multiplying the Pearl potential
by a factor $f(x,y)$ that vanishes at the film edges with
a $\Lambda$ dependent slope and that looks like a conical
mountain whose maximum height for $\Lambda=0$ reaches unity
and for $\Lambda >0$ is somewhat lower than unity. This
numerical finding means that for $\Lambda >0$ the correct,
shape-dependent interaction is smaller than the ideal
Pearl potential. The physical reason for this is that for
finite films the magnetic field lines of a vortex
do not extend to infinity but return around the film edges,
thereby reducing the sheet current that causes the
vortex--vortex interaction, see discussion before Eq.~(1).

  Two examples for this factor
$f(x,y) = V_{\rm num}(x,y) / V_{\rm Pearl}(x,y)$ , i.e.,
the ratio of the exact numerical potential and the ideal
Pearl potential, are depicted in Fig.~6 for a superconducting
thin film square with the first vortex positioned at the
center ($0,0$) and the second vortex at $(x,y)$. This ratio
has approximately conical shape, i.e., the correction factor
$V_{\rm num} / V_{\rm Pearl}$  nearly linearly goes from
unity to zero as the second vortex moves from the center
to the edge of the square. It can be nicely seen in
Fig.~6 that near the edges $V_{\rm num}(x,y)$ vanishes
$\propto \delta^{1/2}$ when $\Lambda=0$ and $\propto \delta$
when $\Lambda > 0$. These plots also show the somewhat
unexpected result that the deviation of the exact
interaction potential from the Pearl potential can be
considerable even when both vortices are far from the
film edges.

 \section{Summary}   

It is shown that the interaction potential between two Pearl
vortices in a superconducting thin {\it infinite} film can be well
approximated by a simple logarithm at all distances,
Eq.~(6) and Fig.~1. A
general argument is given why this interaction at large distances
$r \gg \Lambda$ has the universal behavior $V(r) = \Phi_0^2
/(\mu_0 \pi r)$, independent of the 2D magnetic penetration depth
$\Lambda = \lambda^2 /d$. Explicit expressions are given for the
interaction of one central Pearl vortex with an infinite regular
lattice of pinned Pearl vortices or with the image vortices in
numerical simulations that use periodic boundary conditions. The
curvature of this interaction (spring constant of the central
vortex) is presented both for this regular rigid lattice, and for
the elastically deformable lattice of Pearl vortices. The
modification of these results by the finite extension of the
vortex lattice in finite films (e.g., disks) is given, see Fig.~2.
Finally, it is shown how the interaction of the vortices in thin
films of {\it finite} size and arbitrary shape can be computed.
This general vortex--vortex interaction potential depends not
only on the distance $r = |{\bf r}_1 - {\bf r}_2|$ of the two
Pearl-like vortices but on both positions ${\bf r}_1$
and ${\bf r}_2$. The interaction in general is not central; e.g.,
the interaction force on a vortex near the film edge acts
perpendicular to this edge no matter where the other vortex is
positioned. Anyway, this non-central interaction is caused by the
other vortex and it vanishes when one of the two vortices
approaches the edge. Examples for this correct interaction are
shown in Figs.~3 and 4. Figure~5 shows the corresponding
radial-symmetric and film-shape independent Pearl potential that,
away from the film edges, exhibits the approximately correct peaks
but does not vanish at the (inner and outer) film edges as the
correct potential does. The correct numerical vortex interaction
in finite-size films everywhere in the film is {\it weaker} than
the ideal, infinite-film vortex interaction potential of Pearl,
except when both vortices are close to each other near the middle
of the film. This is shown in Fig.~6 for a square film with one
vortex at the center.

\acknowledgments

The author acknowledges stimulating discussions with
John R.\ Clem and Dieter Koelle.

\appendix
\section{Elasticity theory}  

  The linear elastic restoring force experienced by a
displaced vortex in a 2D lattice of pinned or pin-free
vortices may also be calculated from the theory of elasticity
of the vortex lattice. Within this 2D problem the vortex
lattice is described by its uniaxial compressional modulus
$c_{11}$ and by its shear modulus $c_{66}$, which is typically
much smaller, $c_{66} \ll c_{11}$. Within London theory one
has for the triangular lattice of parallel Abrikosov
vortices \cite{10,11,12}
 \begin{eqnarray}  
   c_{11}(k) = {\bar B^2 /\mu_0 \over 1+k^2 \lambda^2},~~
   c_{66} = {\bar B \Phi_0 \over 16 \pi \mu_0 \lambda^2} \,,
 \end{eqnarray}
where ${\bf k}=(k_x, k_y)$ is the wave vector of the
periodic displacement field. The $k$ dependence
(dispersion) of $c_{11}(k)$ means that the elasticity is
nonlocal. One has the ratio $c_{11}(0) /c_{66}
 = 8b\kappa^2 =(32\pi /\sqrt3) (\lambda/a)^2 \gg 1 $, with
$b = \bar B /B_{c2}$ and $B_{c2} = \Phi_0/(2\pi \xi^2)$
the upper critical field of the superconductor.
Expressions (A1) are valid when the magnetic fields of
the vortices overlap, i.e., for vortex spacings $a \ll
2\pi \lambda$ equivalent to $1/(2\kappa^2) \ll b \ll 1$.
At very small inductions, approximate expressions may be
obtained by considering only nearest neighbor interaction,
yielding $c_{11} \propto c_{66} \propto \exp(-a/\lambda)$.

  Interestingly, for the 2D lattice of Pearl vortices in
thin films the same moduli $c_{11}$, $c_{66}$, Eq.~(A1),
apply (referred to unit volume). Namely, in the limit of
thin films with thickness $d \ll \lambda$ in the London limit
$\kappa \gg 1$ one has \cite{5} for the moduli per unit area
$d c_{11}(k) = d \bar B^2 /(\mu_0 k^2 \lambda^2) =
 \bar B^2 /(\mu_0 k^2 \Lambda)$ (for $k^2\lambda^2 \gg 1$)
and $d c_{66} = \bar B \Phi_0 /(16 \pi \mu_0 \Lambda)$.
Thus, the below results apply both to the lattice of
parallel Abrikosov vortices with length $d \gg a$ and to
thin films with thickness $d \ll \lambda$.

  We introduce the vortex displacements
${\bf u}_\nu = {\bf u(R}_\nu)$ and their Fourier transforms
${\bf u(k)} = [u_x({\bf k}), u_y({\bf k})]$,
 \begin{eqnarray}  
 {\bf u}_\nu = \!\int_{\!BZ} {d^2 k\over 4\pi^2}\,{\bf u(k)}
  \,e^{i {\bf k R_\nu}} \!, ~
 {\bf u(k)} = \!\sum_\nu\! {{\bf u}_\nu \over n}
  \,e^{-i {\bf k R_\nu}} \!.
 \end{eqnarray}
Here $n=\bar B/\Phi_0$ is the vortex density and the
${\bf k}$ integral is over the first Brillouin zone (BZ) with
area $4\pi^2 n = \pi k_{B}^2$, where
$k_B = (4\pi n)^{1/2} \approx \pi/a$ is the radius of the
circle that may be used to approximate the hexagonal BZ.
At the maximum $k=k_B$ one has $c_{11}(k_B)/c_{66} =4$.
The elastic energy per unit area (hence the factor $d$) is
  \begin{eqnarray}  
  F_{\rm elast} = {d\over 2} \!\int_{\!BZ} {d^2 k\over 4\pi^2}
    \,u_\alpha({\bf k}) \Phi_{\alpha\beta}({\bf k})
      u_\beta^*({\bf k}) \,.
  \end{eqnarray}
In it $\Phi_{\alpha\beta}({\bf k})$ with $\alpha,\beta = x,y$
is the elastic matrix, $\Phi_{\alpha\beta}({\bf k}) =
(c_{11}-c_{66})k_\alpha k_\beta + \delta_{\alpha\beta}
c_{66} k^2 $  with $k^2 = k_x^2 +k_y^2$. Explicitly one has
  \begin{eqnarray}  
  \Phi_{xx} &=& c_{11}k_x^2 + c_{66}k_y^2,~  \nonumber
  \Phi_{yy}  =  c_{66}k_x^2 + c_{11}k_y^2 \, \\
  \Phi_{xy} &=& \Phi_{yx} = (c_{11} -c_{66}) k_x k_y\,,
  \end{eqnarray}
the determinant of $\Phi_{\alpha\beta}({\bf k})$ is
$D({\bf k}) = k^4 c_{11} c_{66}$, and the inverse matrix has
the elements  $(\Phi^{-1})_{xx} = \Phi_{yy}/D$,
$(\Phi^{-1})_{yy} = \Phi_{xx}/D$, and
$(\Phi^{-1})_{xy} = (\Phi^{-1})_{yx} = \Phi_{xy}/D$. We shall
now apply this formalism to two problems.

First, we consider the problem of Eq.~(12): all vortices are
rigidly pinned but the free vortex at the origin is shifted
by a distance $u_0$
along $x$, thus ${\bf u}_\nu = u_0 \delta_{\nu,0} {\bf \hat x}$.
Inserting this into (A2) yields ${\bf u(k)}=(u_0/n){\bf\hat x}$
and the elastic energy (A3) becomes
  \begin{eqnarray}  
  F_{\rm elast} = {u_0^2 \over 2}\, V''_{c,\infty}(0) =
  {d\over 2} \int_{\!BZ} {d^2 k\over 4\pi^2} \,
   \Phi_{xx}({\bf k}) \, {u_0^2 \over n^2}\,.
  \end{eqnarray}
Since $c_{11} \gg c_{66}$ one has from Eqs.~(A1) and (A4)
$\Phi_{xx} \approx c_{11}(k)k_x^2 \approx (\bar B^2/\mu_0)
k_x^2 /(1+k^2\lambda^2) \approx (\bar B^2/\mu_0 \lambda^2)
k_x^2 /k^2 $. Averaging $k_x^2 /k^2 = \cos^2 \varphi$ over
the angle $\varphi$ yields $1\over2$. With the BZ area $4\pi^2n$
we thus obtain the potential curvature
  \begin{eqnarray}  
  V''_{c,\infty}(0) = {\bar B^2 d \over \mu_0 \lambda^2 n^2}
  \int_{\!BZ} {d^2 k\over 4\pi^2} \, {k_x^2 \over k^2}
  = {\bar B \Phi_0 d \over 2 \mu_0 \lambda^2} \,.
  \end{eqnarray}
This coincides with Eq.~(14).

Second, we consider the case when all vortices in a cylinder
or disk with radius $R_d$ can move freely, only the vortices
at the surface or edge $r =R_d$ are pinned to ensure zero
displacement there. We calculate the elastic force $f_0$ that
is required to move the central vortex a distance $u_0$,
thereby deforming the vortex lattice. Thus,
the applied forces on the vortices are all zero except for
the force acting on the central vortex,
${\bf f}_\nu = f_0 \delta_{\nu,0} {\bf \hat x}$.
In general, we define the forces ${\bf f}_\nu$ per unit length
(the total force on the $\nu$th vortex of length $d$ is
$d{\bf f}_\nu$) and their Fourier transform
${\bf f(k)} = [f_x({\bf k}), f_y({\bf k})]$ by
 \begin{eqnarray}  
 {\bf f}_\nu = \!\int_{\!BZ} {d^2 k\over 4\pi^2}\,{\bf f(k)}
  \,e^{i {\bf k R_\nu}} \!, ~
 {\bf f(k)} = \!\sum_\nu\! {{\bf f}_\nu \over n}
  \,e^{-i {\bf k R_\nu}} \!.
 \end{eqnarray}
The elastic energy (A3) may also be written as
  \begin{eqnarray}  
  F_{\rm elast}\!\! &=& {d \over 2} \, \sum_\nu \,
    {\bf f}_\nu \, {\bf u}_\nu  \nonumber \\
    &=& {d \over 2}\, n  \int_{\!BZ} {d^2 k\over 4\pi^2}
     \,f_\alpha({\bf k}) \, u_\alpha^*({\bf k}) \nonumber \\
    &=& {d \over 2}\, n^2 \!\!\int_{\!BZ} {d^2 k\over 4\pi^2}
     \,f_\alpha({\bf k}) \,\Phi^{-1}_{\alpha\beta}({\bf k})
       f_\beta^*({\bf k}) \,.
  \end{eqnarray}
Displacements and forces are related by
  \begin{eqnarray}  
    n f_\alpha({\bf k}) =\Phi_{\alpha\beta}({\bf k})
      u_\beta^*({\bf k}) \,, \nonumber \\
    u_\alpha({\bf k}) =n \Phi^{-1}_{\alpha\beta}({\bf k})
      f_\beta^*({\bf k}) \,.
  \end{eqnarray}
For our example
${\bf f}_\nu = f_0 \delta_{\nu,0} {\bf \hat x}$ one has
${\bf f(k)} = (f_0/n) {\bf \hat x}$ and the elastic energy
becomes, with $\Phi^{-1}_{xx} \approx k_y^2/(k^4 c_{66})$,
  \begin{eqnarray}  
  F_{\rm elast} = {d f_0^2 \over 2} \!
   \int_{\!BZ} {d^2 k\over 4\pi^2} \,\Phi_{xx}^{-1}({\bf k})
   = {d f_0^2 \ln(R_d k_B) \over 8 \pi c_{66}} \,.
  \end{eqnarray}
For the ${\bf k}$ integral we have used a lower cut-off
$R_d^{-1}$ and the upper limit $k_B \approx \pi/a$.
The curvature of the elastic potential, $V''_{\rm elast}(0)$,
that the vortex at the origin feels is defined by
  \begin{eqnarray}  
  F_{\rm elast} = {1 \over 2}\, u_0^2 V''_{\rm elast}(0)
  = {1\over 2}\, {d^2 f_0^2 \over V''_{\rm elast}(0)} \,,
  \end{eqnarray}
since the central force is $d f_0 =V''_{\rm elast}(0) u_0$,
i.e., $V''_{\rm elast}(0)$ is a spring constant. Comparing
Eqs.~(A10) and (A11) and using the shear modulus (A1)
we obtain
  \begin{eqnarray}  
  V''_{\rm elast}(0) ={\bar B \, \Phi_0\, d \over
  4\mu_0\lambda^2 \ln(R_d k_B)}
  ={V''_{c,\infty}(0) \over2 \ln(\pi R_d/a)}\,.
  \end{eqnarray}
Thus, when the surrounding vortices are allowed to relax
elastically, the displacement of the central vortex, on which
a force acts, is increased by a factor $2 \ln(\pi R_d/a)$.


\end{document}